\pdfoutput=1
\documentclass{article}
\usepackage{cite}
\usepackage{spconf,amsmath,graphicx}
\usepackage{algorithmic}
\usepackage{multirow}
\usepackage{threeparttable}
\usepackage{enumitem}
\usepackage{subfigure}
\usepackage{makecell}
\usepackage{caption}
\usepackage{xcolor}
\title{Rethinking complex-valued deep neural networks for monaural speech enhancement}

\name{Haibin Wu, Ke Tan, Buye Xu, Anurag Kumar, Daniel Wong
\thanks{This work was done while H. Wu was a research scientist intern at Meta.}
}

\address{
Meta Reality Labs Research, USA
}

\begin{document}
\ninept

\maketitle
\ninept

\begin{abstract}
Despite multiple efforts made towards adopting complex-valued deep neural networks (DNNs), it remains an open question whether complex-valued DNNs are generally more effective than real-valued DNNs for monaural speech enhancement. 
This work is devoted to presenting a critical assessment by systematically examining complex-valued DNNs against their real-valued counterparts.
Specifically, we investigate complex-valued DNN atomic units, including linear layers, convolutional layers, long short-term memory (LSTM), and gated linear units.
By comparing complex- and real-valued versions of fundamental building blocks in the recently developed gated convolutional recurrent network (GCRN), we show how different mechanisms for basic blocks affect the performance.
We also find that the use of complex-valued operations hinders the model capacity when the model size is small.
In addition, we examine two recent complex-valued DNNs, i.e. deep complex convolutional recurrent network (DCCRN) and deep complex U-Net (DCUNET). Evaluation results show that both DNNs produce identical performance to their real-valued counterparts while requiring much more computation.
Based on these comprehensive comparisons, we conclude that complex-valued DNNs do not provide a performance gain over their real-valued counterparts for monaural speech enhancement, and thus are less desirable due to their higher computational costs.

\end{abstract}

\section{Introduction}
\label{sec:intro}

Recent years have witnessed promising performance improvement of monaural speech enhancement models in the complex domain, given the importance of phase for speech quality~\cite{williamson2015complex,fu2017complex,choi2018phase,tan2019complex,tan2019learning,ouyang2019fully,hu2020dccrn,xu2020listening,li2021two}.
A recent study \cite{Chiheb17deepcomplex} develops the key atomic components for complex-valued DNNs and claim that complex-valued parameters have various merits from computational, biological, and signal processing perspectives.
Complex-valued DNNs, which operates with complex-valued arithmetic, seems to be advantageous for complex-domain speech enhancement, where DNNs are trained to learn complex spectrograms.
Motivated by such an intuition, multiple efforts \cite{hu2020dccrn,choi2018phase,lv2021dccrn+,lv2022s,zhao2021monaural,sun2021funnel,watcharasupat2022end,zou2021dctcn} adopted complex-valued DNNs for monaural speech enhancement.
However, to the best of our knowledge, none of these studies has justified a performance gain provided by complex-valued DNNs over their real-valued counterparts with the same network structure and model size. Drude et al. \cite{drude2016appropriateness} compared real- and complex-valued DNNs with fully-connected layers for beamforming, and found that the complex-valued DNN does not yield superior performance to the real-valued DNN while more computationally expensive. For monaural speech enhancement, despite the promising performance improvement produced by recent complex-valued DNNs, it remains unclear whether it is the complex-valued nature that fundamentally brings the merits.

A recent notable model named DCCRN \cite{hu2020dccrn} extends the convolutional recurrent network in~\cite{tan2018convolutional} by replacing convolutional and LSTM layers with their complex-valued counterparts to estimate the ideal complex ratio mask. The DCCRN exhibits competitive performance over earlier works, which has drawn the community's attention to the efficacy of complex-valued DNNs for speech enhancement.
However, we believe that it is premature to ascribe the performance improvement to the use of complex-valued operations due to the lack of systematic comparisons between DCCRN and its real-valued counterpart, in which only the complex-valued layers are replaced by the corresponding real-valued layers while all other configurations remain unaltered, including input features, training targets, training objectives, network structure and model size.  Without such apples-to-apples comparisons, it is difficult to justify the attribution of the improvement achieved by complex-valued DNNs.

This study presents a critical assessment by systematically examining complex-valued DNNs against their real-valued counterparts through comprehensive comparisons:

\noindent 1). Based on the principles of complex-valued computation \cite{Chiheb17deepcomplex}, we formulate complex-valued DNN atomic units for investigation, including linear layers, convolutional/deconvolutional layers, LSTM, and gated linear units. 
We compare their performance with that of their real-valued counterparts on monaural speech enhancement.

\noindent 2). To comprehensively investigate complex-valued operations in different types of layer topology, we adopt GCRN - a real-valued DNN originally developed for complex-domain speech enhancement, which integrates a variety of layer types. We enumerate all the different versions of fundamental building blocks of GCRN, and show how different computing mechanisms in basic blocks affect the performance. We observe that the models with complex-valued components do not outperform the real-valued counterparts. In addition, given the fact that many real-world applications require a computationally efficient model, we conduct the same comparisons with a setting where the model size is very small. We find that, with such a setting, complex-valued operations even hinders speech enhancement performance compared to real-valued operations.

\noindent 3). Two recent compelling models based on complex-valued operations, DCCRN \cite{hu2020dccrn} and DCUNET \cite{choi2018phase}, have shown promising performance for monaural speech enhancement. In this work, we evaluate their real-valued versions with the same parameter count, and conduct investigation with different loss functions, learning rates and minibatch sizes, in terms of both enhancement performance and training stability. The experimental results reveal that the complex-valued versions do not outperform their real-valued counterparts while they have higher computational costs. This is consistent with the observation in~\cite{pandey2019exploring}.

\section{Methodology}
\label{sec:method}
This section introduces the basic building blocks for complex-valued DNNs, followed by the case study design.

\subsection{Building blocks}
\label{sec:building blocks}
\subsubsection{Linearity}
\label{sec:method:linearity}
Fully connected layers, convolution layers and deconvolution layers are composed of matrix multiplications. 
We omit the bias to simplify the description.
Taking the input complex-valued feature matrix as $X=X_{r} + jX_{i}$ and the complex-valued parameter matrix as $W=W_{r} + jW_{i}$, the complex-valued output can be elaborated as:
\begin{equation}
\label{eq:matrix multiply}
\begin{aligned}
Y = (X_{r} W_{r}-X_{i} W_{i})+j(X_{r} W_{i}+X_{i} W_{r}),
\end{aligned}
\end{equation}
where $Y$ denotes the output feature of the complex-valued layer, the subscripts $r$ and $i$ denote real and imaginary parts respectively.

\subsubsection{Activation function}
\label{sec:method:activation}
Given a complex-valued representation $z$, the activation function operates on the real and imaginary part independently as:
\begin{equation}
\label{eq:activation function}
\begin{aligned}
a = f(Re\ z) + j f(Im\ z),
\end{aligned}
\end{equation}
where $a$ is the output representation, $Re$ and $Im$ extract the real and imaginary parts respectively, and $f$ denotes the activation function.

\subsubsection{LSTM}
\label{sec:lstm}
For LSTM layers, we have two versions:

\noindent
\textbf{Quasi complex-valued LSTM} 
In \cite{hu2020dccrn}, the complex LSTM operation is treated as two separate operations on the real and imaginary parts. To be specific, they initialize two real-valued sub-LSTM layers, namely $\text{LSTM}_{r}$ and $\text{LSTM}_{i}$, corresponding to the real and imaginary LSTM respectively. Given the input feature $X=X_{r} + jX_{i}$, the output feature can be derived as:
\begin{equation}
\label{eq:DCCRN-style LSTM}
\begin{aligned}
& F_{rr} = \text{LSTM}_{r}(X_{r}), F_{ir} = \text{LSTM}_{r}(X_{i}), \\
& F_{ri} = \text{LSTM}_{i}(X_{r}), F_{ii} = \text{LSTM}_{i}(X_{i}), \\
& F_{out} = (F_{rr}-F_{ii}) + j(F_{ri} + F_{ir}),
\end{aligned}
\end{equation}
where $F_{out}$ is the output feature.

\noindent
\textbf{Fully complex-valued LSTM} In addition to the quasi complex-valued LSTM, which does not perform complex-valued operations within sub-LSTM layers, we also investigate fully complex-valued LSTM, which totally follows the the arithmetic of complex numbers.
Each matrix multiplication and activation function in this LSTM strictly follows the arithmetic in Sections~\ref{sec:method:linearity} and~\ref{sec:method:activation}.

\subsubsection{Gated linear unit}
\label{sec:gating}
Gated linear unit~\cite{dauphin2017language} is a widely used layer topology, which consists of two separate convolutional layers and one gating operation.
The two separate convolutional layers process the same input, and produce their outputs $F^{(1)}$ and $F^{(2)}$, respectively. A sigmoid function is applied to $F^{(2)}$ to dervie a gate, which is then element-wisely multiplied with $F^{(1)}$ to yield the output of the gated linear unit.
In a complex-valued gated linear unit, let $F^{(1)}=F_{r}^{(1)} + jF_{i}^{(1)}$ and $F^{(2)}=F_{r}^{(2)} + jF_{i}^{(2)}$ be the outputs of the two convolutional layers. 
We derive two gating mechanisms, i.e. separate gating and magnitude gating.

\noindent
\textbf{Separate gating} For separate gating, we apply a sigmoid function to the real and imaginary parts of $F^{(2)}$ separately, which amounts to a complex-valued gate. The real and imaginary parts of this gate are element-wisely multiplied with $F_{r}^{(1)}$ and $F_{i}^{(1)}$, respectively.

\noindent
\textbf{Magnitude gating} Unlike separate gating, magnitude gating calculates a real-valued gate $F^{(g)}$ from the magnitude of the complex feature map $F^{(2)}$:
\begin{equation}
\label{eq:magnitude gating}
\begin{aligned}
F^{(g)}=(\sigma(|F^{(2)}|) - 0.5) \times 2,
\end{aligned}
\end{equation}
where $\sigma$ denotes the sigmoid function, and $|\cdot|$ extracts the magnitude of a complex feature map.
Since the magnitude is nonnegative, applying the sigmoid function to the magnitude always results in values ranging from 0.5 to 1. 
Hence we use an affine transformation to normalize the gating value to the range of 0 to 1. The resulting gate is applied to both real and imaginary parts of $F^{(1)}$.
Such magnitude gating preserves the phase of $F^{(1)}$~\cite{hayakawa2018applying}.


\begin{table*}[h]
\footnotesize
\caption{Investigation of different basic units, where the number of multiply-accumulate (MAC) operations is measured on a 1-second signal.}
\vspace{-5pt}
\centering
\resizebox{0.9\textwidth}{!}{
\begin{tabular}{lc|c|ccc|cc|cc}
\hline
\multirow{2}{*}{}                                         
& \multirow{2}{*}{SNR} 
& \multirow{2}{*}{Noisy} & \multirow{2}{*}{(1a).C-LSTM}& \multirow{2}{*}{(1b).Quasi C-LSTM} & \multirow{2}{*}{(1c).LSTM} & \multirow{2}{*}{(1d).C-Linear} & \multirow{2}{*}{(1e).R-Linear} & \multirow{2}{*}{(1f).DCUNET} & \multirow{2}{*}{(1g).RUNET}
\\ & & & & & & & \\ \hline
\multirow{3}{*}{\begin{tabular}[c]{@{}c@{}}STOI\end{tabular}} 
&-5 dB &0.69 &0.85	&0.86 &0.86	&0.61	&0.61	&0.85	&0.85 \\
&0 dB  &0.78 &0.90	&0.91 &0.91	&0.70	&0.70	&0.90	&0.90 \\
&5 dB &0.85	 &0.94	&0.94 &0.94	&0.76	&0.76	&0.94	&0.94 \\

\hline
\multirow{3}{*}{\begin{tabular}[c]{@{}c@{}}WB-PESQ\end{tabular}} 
&-5 dB	&1.11	&1.65	&1.71 &1.69	&1.12	&1.12	&1.64	&1.70 \\
&0 dB	&1.15	&1.95	&2.02 &2.00	&1.17	&1.28	&1.92	&2.00 \\
&5 dB	&1.24	&2.29	&2.35 &2.34	&1.24	&1.25	&2.27	&2.36 \\
\hline
\multirow{3}{*}{\begin{tabular}[c]{@{}c@{}}SI-SDR (dB)\end{tabular}} 
&-5 dB &-5.00 &10.80	&11.10 &10.87	&0.92	&1.23	&10.80	&10.87 \\
&0 dB &0.05	  &13.62	&13.92 &13.78	&4.69	&4.94	&13.79	&13.86 \\
&5 dB &5.01	  &16.36	&16.64 &16.55	&7.19	&7.57	&16.74	&16.84 \\
\hline
$\#$ Para & - & -	&23.35 M	&23.35 M &23.62 M	&0.59 M	&0.59 M	&3.10 M	&3.12 M
\\ \hline
$\#$ MACs & - & -	&5.90 G	&5.90 G &2.98 G	&119.59 M	&59.88 M	&56.69 G	&19.87 G
\\ \hline
\end{tabular}
}
\label{tab:basic units}
\end{table*}

\begin{table*}[h]
\vspace{-5pt}
\scriptsize
\centering
\caption{Investigation of different complex-valued components in GCRN. $\clubsuit$, $\diamondsuit$, $\heartsuit$ denote using the quasi complex-valued LSTM in the bottleneck, complex-valued convolutional layers, complex-valued deconvolutional layers, respectively. ``- Separate'' and ``- Magnitude'' denote using separate and magnitude gating mechanisms in GLUs, respectively, and $\odot$ denotes the model performs complex ratio masking rather than complex spectral mapping originally used in~\cite{tan2019learning}.}
\centering
\vspace{-5pt}
\resizebox{0.9\textwidth}{!}{
\begin{tabular}{cl|ccc|ccc|ccc|cc}
\hline
\multicolumn{1}{c}{\multirow{2}{*}{}} & & \multicolumn{3}{c|}{\textbf{STOI}} & \multicolumn{3}{c|}{\textbf{WB-PESQ}} & \multicolumn{3}{c|}{\textbf{SI-SDR (dB)}} & \multirow{2}{*}{\textbf{\# Para}} & \multirow{2}{*}{\textbf{\# MACs}}\\
\multicolumn{1}{c}{}    &     &-5 dB	&0 dB	&5 dB	&-5 dB	&0 dB	&5 dB	&-5 dB	&0 dB	&5 dB
\\ \hline
 & Noisy	&0.69	&0.78	&0.85	&1.11	&1.15	&1.24	&-5	&0.01	&5.01 & - & -        
\\ \hline
(2a) & GCRN (real-valued model) & 0.84 & 0.90 & 0.94	& 1.57 & 1.87 & 2.24 & 8.30 & 11.29 & 14.13 & 9.25 M & 1.72 G 
\\ 
(2b) & GCRN + $\clubsuit$	& 0.83 & 0.90 & 0.94 & 1.55 & 1.85 & 2.22	& 8.22 & 11.17 & 13.98 & 9.25 M & 2.57 G
\\
(2c) & GCRN + $\diamondsuit$ - Separate	& 0.83 & 0.90 & 0.93 & 1.53 & 1.80 & 2.15 & 7.64 & 10.51 & 13.22 & 9.12 M & 1.72 G 
\\
(2d) & GCRN + $\diamondsuit$ - Magnitude & 0.83 & 0.90 & 0.94 & 1.56 & 1.85 & 2.23 & 7.66 & 10.63 & 13.43 & 9.12 M & 1.72 G            
\\
(2e) & GCRN + $\clubsuit$ + $\diamondsuit$ - Separate & 0.83 & 0.90 & 0.93 & 1.52 & 1.81 & 2.16 & 8.14 & 11.16 & 14.02 & 9.12 M & 2.57 G
\\
(2f) & GCRN + $\clubsuit$ + $\diamondsuit$ - Magnitude	& 0.84 & 0.90 & 0.94 & 1.56 & 1.87 & 2.24 & 7.89 & 10.89 & 13.76 & 9.12 M & 2.57 G
\\
(2g) & GCRN + $\diamondsuit$ + $\heartsuit$ - Separate	&0.83	&0.89	&0.93	&1.53	&1.83	&2.20	&7.95	&10.96	&13.88	&8.83 M	&1.72 G
\\
(2h) & GCRN + $\diamondsuit$ + $\heartsuit$ - Magnitude	&0.83	&0.90	&0.94	&1.54	&1.85	&2.23	&7.67	&10.75	&13.79	&8.83 M	&1.72 G
\\
(2i) & GCRN + $\clubsuit$ + $\diamondsuit$ + $\heartsuit$ - Separate	&0.82	&0.89	&0.93	&1.52	&1.80	&2.15	&7.62	&10.70	&13.52	&8.83 M	&2.57 G
\\
(2j) & GCRN + $\clubsuit$ + $\diamondsuit$ + $\heartsuit$ - Magnitude	&0.83	&0.90	&0.94	&1.57	&1.88	&2.27	&7.65	&10.87	&13.82	&8.83 M	&2.57 G
\\
(2A) & GCRN $\odot$	(real-valued model) & 0.83 & 0.89 & 0.93 & 1.50 & 1.79 & 2.16 & 7.28 & 10.33 & 13.40 & 9.25 M & 1.72 G \\
(2J) & GCRN + $\clubsuit$ + $\diamondsuit$ + $\heartsuit$ - Magnitude $\odot$	&0.82	&0.89	&0.93	&1.47	&1.74	&2.10	&7.25	&10.33	&13.46	&8.83 M	&2.57 G
\\
\hline

\end{tabular}
}
\label{tab:GCRN}
\vspace{-10pt}
\end{table*}

\begin{table}[h]
\tiny
\caption{Comparison between GCRN and CGCRN with relatively small model sizes. Subscripts ``M'' and ``S'' denote a middle size and a small size, respectively.}
\vspace{-5pt}
\centering
\resizebox{0.48\textwidth}{!}{
\begin{tabular}{lc|c|cc|cc}
\hline
\multirow{2}{*}{}                                         
& \multirow{2}{*}{SNR} & \multirow{2}{*}{Noisy} & \multirow{2}{*}{CGCRN$_\text{M}$} & \multirow{2}{*}{GCRN$_\text{M}$}& \multirow{2}{*}{CGCRN$_\text{S}$} & \multirow{2}{*}{GCRN$_\text{S}$}
\\ & & & & \\ \hline
\multirow{3}{*}{\begin{tabular}[c]{@{}c@{}}STOI\end{tabular}} 
&-5 dB	&0.69 &0.81	& 0.81	&0.75	&0.79 \\
&0 dB	&0.78 &0.88	& 0.88 &0.83	&0.86             \\
&5 dB	&0.85 &0.92	& 0.92 &0.88	&0.91  \\ \hline
\multirow{3}{*}{\begin{tabular}[c]{@{}c@{}}WB-PESQ\end{tabular}} 
&-5 dB	&1.11 &1.39	& 1.42 &1.29	&1.35  \\
&0 dB	&1.15 &1.62	& 1.64 &1.47	&1.54   \\
&5 dB	&1.24 &1.93	& 1.95 &1.71	&1.82          \\ \hline
\multirow{3}{*}{\begin{tabular}[c]{@{}c@{}}SI-SDR (dB)\end{tabular}} 
&-5 dB	&-5.00 & 6.60 &	7.25 & 4.14	& 5.82           \\
&0 dB	&0.05 &9.75	& 10.26 & 6.80	& 8.94       \\
&5 dB	&5.01 &12.58 & 12.94 & 8.73	&11.85      \\ 
\hline
$\#$ Para (M) & - & - & 2.26 & 2.36 & 0.61	& 0.63 
\\ \hline
$\#$ MACs (M) & - & - &	657.90 & 439.03 & 172.12 & 115.99
\\ \hline
\end{tabular}
}
\label{tab:tiny GCRN}
\vspace{-5pt}
\end{table}

\begin{table*}[h]
\caption{Comparisons between real- and complex-valued versions of DCCRN with different training objectives. ``-Real'' means the real-valued version of DCCRN. ``-SISDR'', ``-L$_\text{1}$'', ``-MSE'' denote using SI-SDR, L$_\text{1}$ and mean squared error (MSE) losses for training, respectively, where both L$_\text{1}$ and MSE losses are computed on the clean and estimated real, imaginary and magnitude spectrograms.}
\vspace{-5pt}
\centering
\resizebox{0.9\textwidth}{!}{
\begin{tabular}{lc|c|cc|cc|cc}
\hline
\multirow{2}{*}{}                                         
& \multirow{2}{*}{SNR} & \multirow{2}{*}{Noisy} 
& \multirow{2}{*}{DCCRN-SISDR} & \multirow{2}{*}{DCCRN-Real-SISDR} & \multirow{2}{*}{DCCRN-L$_\text{1}$}& \multirow{2}{*}{DCCRN-Real-L$_\text{1}$} & \multirow{2}{*}{DCCRN-MSE} & \multirow{2}{*}{DCCRN-Real-MSE}
\\ & & & & & & \\ \hline
\multirow{3}{*}{\begin{tabular}[c]{@{}c@{}}STOI\end{tabular}} 
&-5 dB &0.69 &0.87	&0.87	&0.86	&0.86	&0.85	&0.85 \\
&0 dB  &0.78 &0.92	&0.92	&0.91	&0.91	&0.90	&0.90 \\
&5 dB  &0.85 &0.95	&0.95	&0.95	&0.95	&0.94	&0.94 \\
\hline
\multirow{3}{*}{\begin{tabular}[c]{@{}c@{}}WB-PESQ\end{tabular}} 
&-5 dB &1.11 &1.78	&1.80	&1.73	&1.69	&1.55	&1.56 \\
&0 dB  &1.15 &2.13	&2.14	&2.05	&2.00	&1.83	&1.86 \\
&5 dB  &1.24 &2.51	&2.54	&2.43	&2.38	&2.16	&2.19 \\
\hline
\multirow{3}{*}{\begin{tabular}[c]{@{}c@{}}SI-SDR (dB)\end{tabular}} 
&-5 dB &-5.00 &11.01	&11.06	&8.36	&8.28	&8.09	&8.18 \\
&0 dB  &0.05 &14.00	&14.06	&11.25	&11.19	&11.20	&11.27 \\
&5 dB  &5.01 &16.99	&17.05	&14.36	&14.27	&14.41	&14.54 \\
\hline
\end{tabular}
}
\label{tab:DCCRN different loss}
\end{table*}

\begin{table*}[h]
\caption{Comparisons between real- and complex-valued versions of DCCRN with different training objectives on the DNS Challenge synthetic test set without reverberation.}
\vspace{-5pt}
\centering
\resizebox{0.9\textwidth}{!}{
\begin{tabular}{l|c|cc|cc|cc}
\hline     
\multirow{2}{*}{}  & \multirow{2}{*}{Noisy} & \multirow{2}{*}{DCCRN-SISDR} & \multirow{2}{*}{DCCRN-Real-SISDR}& \multirow{2}{*}{DCCRN-L$_\text{1}$} & \multirow{2}{*}{DCCRN-Real-L$_\text{1}$} & \multirow{2}{*}{DCCRN-MSE} & \multirow{2}{*}{DCCRN-Real-MSE}
\\ & & & & & \\
\hline
STOI &0.92	&0.97	&0.97	&0.97	&0.97	&0.97	&0.97 \\
WB-PESQ &1.58	&2.92	&2.89	&2.92	&2.86	&2.61	&2.64 \\
SI-SDR (dB) &9.23	&19.60	&19.54	&17.11	&17.13	&17.33	&17.55 \\
DNSMOS (OVRL) & 2.48 & 3.30 & 3.33 & 3.28 & 3.30 & 3.19 & 3.20 \\
NORESQA-MOS & 1.90 & 4.31 & 4.34 & 4.27 & 4.31 & 3.80 & 3.96 \\
\hline
$\#$ Para & - &3.67 M	&3.64 M	&3.67 M	&3.64 M	&3.67 M	&3.64 M
\\ \hline
$\#$ MACs & - &14.38 G	&4.84 G	&14.38 G	&4.84 G	&14.38 G	&4.84 G
\\ \hline
\end{tabular}
}
\label{tab:DNS testing set}
\vspace{-10pt}
\end{table*}

\subsection{Case study design}
In this section, we carefully design the case studies, and elaborate the rationales and objectives of each case study. In these case studies, all pairs of real- and complex-valued models for comparison have the same configurations, including input features, training targets, training objectives, network structure and model size.

\noindent
\textbf{Basic Unit} 
This case study compares different complex layers defined in Section~\ref{sec:building blocks} with their real-valued counterparts, in terms of enhancement performance and computational costs.
Specifically, we compare:
1) a model with a stack of three complex-valued linear layers and its corresponding real-valued model, where each of the two hidden layers has 406 units in the complex-valued model and 512 units in the real-valued model, respectively. Such a configuration ensures that the two models have almost the same number of parameters. Note that each hidden layer is followed by a rectified linear unit function; 2) quasi complex-valued LSTM, fully complex-valued LSTM, and real-valued LSTM, each of which contains three LSTM layers followed by a linear output layer. In these three models, each LSTM layer contains 732, 732 and 1024 units, respectively. The implementations described in Section~\ref{sec:lstm} are adopted for quasi complex-valued LSTM and fully complex-valued LSTM; 3) DCUNET, a convolutional encoder-decoder model developed in~\cite{choi2018phase}, and its real-valued counterpart (RUNET), in which all complex-valued convolutional, deconvolutional and linear layers are replaced by their real-valued counterparts. Akin to 1) and 2), we slightly adjust hyperparameters (e.g. number of out channels in convolutional layers) for RUNET, such that its model size is almost the same as DCUNET. Note that all these models are trained to learn complex spectral mapping.

\noindent
\textbf{GCRN} GCRN \cite{tan2019learning} is a representative model for our investigation, because it consists of different types of layers including convolutional/deconvolutional layers, gated linear units, LSTM layers, and linear layers. The original GCRN has two decoders, one for real part estimation and the other for imaginary part estimation. We instead use a single shared decoder for both real and imaginary parts, corresponding to two output channels in the last deconvolutional layer of the decoder. Such an architecture can be naturally converted into complex-valued versions for comparison by replacing each layer with their complex-valued counterpart. In this case study, we aim to investigate:
1) whether replacing specific layers of GCRN with their complex-valued counterparts can lead to better performance;
2) how the use of complex-valued operations affect speech enhancement performance when the model is constrained to a relatively small amount of parameters;
3) which gating mechanism in Section~\ref{sec:gating} is the better choice, from both training stability and enhancement performance aspects.
Note that regarding the bottleneck LSTM in GCRN, we adopt the quasi complex-valued LSTM for investigation.

\noindent
\textbf{DCCRN} In~\cite{hu2020dccrn}, the performance gain achieved by DCCRN is attributed by the authors to the complex multiplication constraint, which they believe can help DNNs learn complex representations more effectively.
However, they did not compare DCCRN with its real-valued counterpart using the same configurations.
Thus it is difficult to justify the attribution of the performance improvement, which is likely due to either the use of complex-valued operations or other components in the model design.
The objective of this case study is to show whether DCCRN can outperform its real-valued counterparts, with the same amount of parameters.
Specifically, we adopt the ``DCCRN-E" configuration, which achieves the best performance in~\cite{hu2020dccrn}.
To derive the corresponding real-valued version, we simply replace the complex-valued layers with their real-valued counterparts, and reduce the channel numbers in the encoder to [32, 64, 64, 64, 128, 256] to maintain the number of parameters.


\section{Experiments}
\label{sec:expt}

\subsection{Experimental setup}
\label{subsec:exp setup}

In our experiments, the Interspeech2020 DNS Challenge training speech dataset~\cite{reddy2020interspeech} is used to create our training, validation and test sets, which contains roughly 65000 speech signals uttered by 1948 speakers in total. We randomly split these speakers into three distinct sets for training, validation and test sets, which include 1753 ($\sim$90\%), 97 ($\sim$5\%) and 98 ($\sim$5\%) speakers, respectively. Similarly, we partition the DNS Challenge noise dataset with around 65000 signals into 90\%, 5\% and 5\% for training, validation and test sets, respectively. By randomly pairing speech and noise signals, we create a training set with 500000 noisy mixtures and a validation set with 1000 noisy mixtures, in both of which the signal-to-noise ratio (SNR) is randomly sampled between -5 and 5 dB. Following the same procedure, three test sets are created at different SNR levels, i.e. -5, 0 and 5 dB. Note that all speech and noise signals are randomly truncated to 10 seconds before mixing. We additionally use the synthetic test set released by DNS Challenge for evaluation.

All signals are sampled at 16~kHz. Short-time Fourier transform is performed to obtain spectrograms. We adopt the Adam optimizer to train all models. Multiple metrics are employed to measure the speech enhancement performance, including wide-band perceptual evaluation speech quality (WB-PESQ) \cite{rec2005p}, short-time objective intelligibility (STOI) \cite{taal2011algorithm}, scale-invariant signal-to-distortion ratio (SI-SDR) \cite{le2019sdr}, DNSMOS P. 835~\cite{reddy2022dnsmos} and NORESQA-MOS~\cite{manocha2022speech}.



\subsection{Experimental results}
\textbf{Basic Unit} In Table~\ref{tab:basic units}, 1). columns (1a), (1b), (1c) denote the fully complex-valued LSTM, quasi complex-valued LSTM and real-valued LSTM. 
2). Real-valued LSTM has half as many MACs as both complex-valued LSTMs.
Among the three models, the quasi complex-valued LSTM achieves the best performance, while its improvement over the real-valued LSTM is marginal.
Columns (1d) and (1e) denote the complex- and real-valued DNNs consisting of linear layers.
Although the real-valued DNN only has half of the MAC number in the complex-valued DNN, it still produces slightly better performance than the latter.
3). (1f) and (1g) denote the DCUNET and its corresponding real-valued version respectively
We see that the real-valued UNET outperforms DCUNET in terms of both enhancement performance and computational efficiency.

\noindent
\textbf{GCRN} In Table~\ref{tab:GCRN}, (2a) is the original real-valued GCRN. (2b)-(2j) are the models where some components are replaced by the corresponding complex-valued version. Moreover, (2A) and (2J) have the same model structure as (2a) and (2j), but are trained to perform complex ratio masking rather than complex spectral mapping.
In Table~\ref{tab:tiny GCRN}, we reduce the model size to roughly 2~M and 0.6~M, where ``CGCRN'' denotes the same configuration as (2j).
We can observe:
1). Replacing the components of GCRN with their complex-valued versions can not get any performance gain, as shown in (2a)-(2j).
2). In the comparison between the models trained for complex ratio masking, i.e. (2A) and (2J), the real-valued model performs slightly better than the complex-valued model.
3). Although the magnitude gating and separate gating lead to similar performance, the training loss curve of the former is much more stable than that of the latter. It is likely because the magnitude gating preserves phase information which could help stabilize the training.
4). In the small model setting, the real-valued models consistently outperforms the complex-valued counterparts.
Furthermore, their performance gap increases as the model size becomes smaller.

\noindent
\textbf{DCCRN} Tables~\ref{tab:DCCRN different loss} and~\ref{tab:DNS testing set} compare the DCCRN with its real-valued counterpart on our simulated test set and the DNS Challenge synthetic test set, respectively.
The following observations are obtained:
1). With three different training objectives, i.e. SI-SDR, L$_1$ and MSE, the real- and complex-valued models yield almost identical performance in all the metrics on both datasets. Take, for example the -5~dB case with the SI-SDR training loss in Table~\ref{tab:DCCRN different loss}. The STOI, WB-PESQ and SI-SDR improvements over noisy mixtures are 0.18, 0.67 and 16.01~dB for the complex-valued model, and 0.18, 0.69 and 16.06~dB for the real-valued model, respectively.
2) As shown in Table~\ref{tab:DNS testing set}, the real-valued model produces slightly better scores than the complex-valued model in both DNSMOS and NORESQA-MOS, i.e. two metrics that highly correlate with subjective quality scores.
3). We have also made comparisons under settings with different learning rates and minibatch sizes. We find that DCCRN is less robust than its real-valued counterpart against different learning rates. In addition, both models produce very similar performance with different minibatch sizes. However, we do not show these comparison results due to the page limit.
4). The real-valued model has only one-third of the MAC amount in the complex-valued model. Specifically, the number of MACs for the complex-valued model is 14.38~G, while it is only 4.84G for the real-valued model. Given that the two models yield almost the same performance, the complex-valued model is less efficient for real-world applications.


\section{Concluding remarks}
\label{sec:conclusion}

Through the extensive experiments, we draw the following conclusions for monaural speech enhancement:
1). Complex-valued DNNs yield similar performance to their real-valued counterparts with the same number of parameters.
2). When the model size is relatively small, the use of complex-valued operations is detrimental to the enhancement performance.
3). The performance gain achieved by DCCRN and DCUNET is not attributed to the use of complex-valued operations. Furthermore, complex-valued DNNs require more MACs than their real-valued counterparts, without any performance gain.

A complex number multiplication can break into four real number multiplications. Based on our systematic comparisons, we believe that real-valued DNNs have the capacity to achieve comparable performance to their complex-valued counterparts with the same model size and network structure. Although complex-valued DNNs intuitively seem a more natural choice than real-valued DNNs for processing complex spectrograms, they are more computationally expensive and thus an inferior choice for real applications that are efficiency-sensitive. We believe that there is no sufficient evidence justifying the superiority of complex-valued DNNs over real-valued DNNs for monaural speech enhancement. This study demonstrates that it is more than nontrivial to rethink the efficacy of complex-valued operations in speech enhancement systems.

\ninept

\bibliographystyle{IEEEbib}
\bibliography{refs}

\end{document}